\def\av#1{\langle#1\rangle}
\documentclass[pre,twocolumn,superscriptaddress,showpacs,%
preprintnumbers,amsmath,amssymb]{revtex4}
\usepackage{amsmath}
\usepackage{graphicx}
\usepackage{epsfig}
\usepackage{dcolumn}
\usepackage{amssymb}    

\begin{document}
\title{Designing optimal transport networks}

\author{G. Li}
\affiliation{Center for Polymer Studies, Department of Physics,
Boston University, 590 Commonwealth Avenue, Boston, USA}
\author{S. D. S. Reis}
\affiliation{Departamento de F\'{\i}sica, Universidade Federal
do Cear\'a, 60451-970 Fortaleza, Cear\'a, Brazil}
\author{A. A. Moreira}
\affiliation{Departamento de F\'{\i}sica, Universidade Federal
do Cear\'a, 60451-970 Fortaleza, Cear\'a, Brazil}
\author{S. Havlin}
\affiliation{Department of Physics, Minerva Center, Bar Ilan University, 
Ramat Gan 52900, Israel}
\author{H. E. Stanley}
\affiliation{Center for Polymer Studies, Department of Physics,
Boston University, 590 Commonwealth Avenue, Boston, USA}
\author{J. S. Andrade Jr.}
\affiliation{Departamento de F\'{\i}sica, Universidade Federal
do Cear\'a, 60451-970 Fortaleza, Cear\'a, Brazil}
\date{\today}
  
\begin{abstract} 

We investigate the optimal design of networks for a general transport
system. Our network is built from a regular two-dimensional ($d=2$)
square lattice to be improved by adding long-range connections
(shortcuts) with probability $P_{ij} \sim r_{ij}^{-\alpha}$, where
$r_{ij}$ is the Euclidean distance between sites $i$ and $j$, and
$\alpha$ is a variable exponent. We introduce a cost constraint on the
total length of the additional links and find optimal transport in the
system for $\alpha=d+1$. Remarkably, this condition remains optimal,
regardless of the strategy used for navigation, being based on local or
global knowledge of the network structure, in sharp contrast with the
results obtained for unconstrained navigation using global or local
information, where the optimal conditions are $\alpha=0$ and $\alpha=d$,
respectively. The validity of our theoretical results is supported by
data on the US airport network, for which $\alpha\approx 3.0$ was
recently found [Bianconi {\it et al.}, arXiv:0810.4412 (2008)].

\end{abstract}

 
\maketitle

The interplay between topology and dynamics in complex systems
represents the focus of many studies in different fields of
research with important scientific and technological applications.
Due to their enormous potential to represent the intricate topology of
numerous systems in nature, complex networks
\cite{Alber1999,Watts1998,Barth1999} have recently been used as
substrates in combination with a plethora of dynamical models to
describe the behavior of biological, social, chemical, physical and
technological networks \cite{Jeong2000,Lawre1999,Giles1998}.
Much attention has been dedicated to the problem of navigation in
complex network geometries 
\cite{Klein2000a,Rober2006,Carmi2009,Cart2009,Komid2008,Mouka2002,Guim2002,Danon2008,Santos2008,Boguna2009}.
In most cases, the influence of the underlying network geography on the
performance of the transport process is investigated assuming that only
local information is available for navigation
\cite{Klein2000a,Rober2006,Carmi2009,Cart2009,Lator2002}.

For many navigation problems of interest in science and technology,
global rather than local information is required, i.e., any source
node $s$ possesses the knowledge of the entire network topology. In
this situation, the average shortest path $\av{\ell}$ from source to
target becomes the relevant navigation variable to be optimized. For
example, in a subway network, such as in Manhattan, the travel routes
should be planned or changed in such a way as to minimize the travel
time for a given limited reconstruction cost. This task is performed
by considering the whole structure of the network in terms of its
nodes and links, namely, by knowing the location of all subway
stations, their connections and the shortest path between any two
stations. If we now consider an underlying network of streets and
avenues over which one has to plan or improve an existing subway
network, and if the aim is to minimize the average travel time between
its stations, the search for an optimal strategy to add new
connections in the network for a given budget should therefore play a
key role. Here we show that the imposition of a cost constraint, which
to the best of our knowledge has not been considered for optimal
navigation, represents a crucial ingredient in the design and
development of efficient navigation networks. 

\begin{figure}[t]
\begin{center}
\includegraphics*[width=6cm]{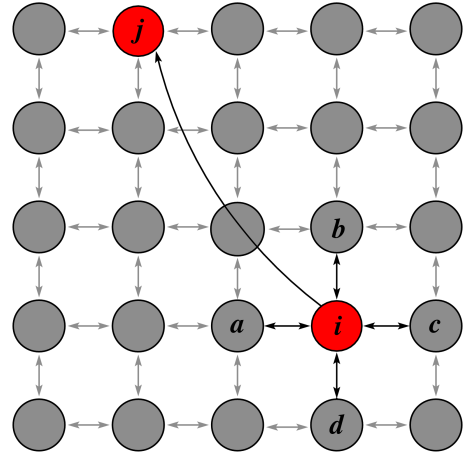}
\end{center}
\caption{(Color online) Connections of a single node $i$. Each 
node $i$ has four short-range connections to its nearest neighbors
($a$, $b$, $c$, and $d$). A long-range directed connection may be
placed to a random chosen node $j$. The node $j$ is selected with
probability proportional to ${r_{ij}}^{-\alpha}$, where $r_{ij}$ is
the distance measured as the number of connections separating the
nodes in the underlying lattice.}
\label{1}
\end{figure}

Consider the case of an existing subway network which needs
improvements \cite{Lator2002,Youn2008}. The financial cost to build up a
large number of {\it new direct connections\/} between distant stations
(i.e., non-neighboring sites) can make it prohibitive, since only
limited resources are normally available for this task. This problem can
be modeled by the following system.  In a $2$-dimensional regular square
lattice, with all $N=L^{2}$ sites present, each site $i$ is connected
with its four nearest neighbors. The sites represent the stations and
the bonds represent the routes of the subway (see Fig.~\ref{1}). In our
model, pairs of sites $ij$ are then randomly chosen to receive
long-range connections with probability proportional to
$r_{ij}^{-\alpha}$, where $r_{ij}$ is the Euclidean distance between
sites $i$ and $j$. Finally, the addition of long-range connections to
the system stops when their total length (cost), $\sum r_{ij}$, reaches
a given value $\Lambda$. Since $\alpha$ controls the average length of
the long-range connections, we obtain that, for a fixed value of
$\Lambda$, and small values of $\alpha$, longer connections, but fewer
in number can be added due to the imposed total length limit. We
therefore expect that an optimal navigation condition must be revealed
as a trade-off between the length and the number of connections added to
the system.

\begin{figure}[t]
\begin{center}
\includegraphics*[width=9cm]{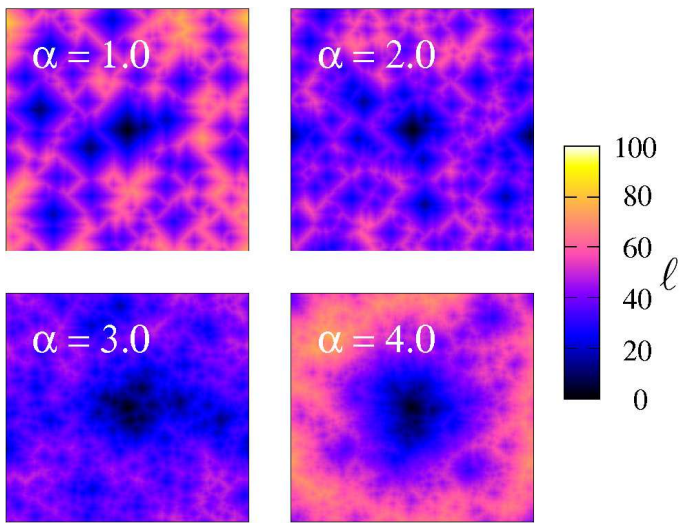}
\end{center}
\caption{(Color online) Shortest-path length, $\ell$, from each node 
to the central node in the network for different values of $\alpha$. In
this case we impose a constraint in the length of the long-range
connections. The sum of the length of these connections is limited,
$\Lambda=\sum r_{ij}=N$, where $N$ is the number of nodes in the
underlying lattice. The network model is constructed from a square
lattice with $L^{2}$ nodes, with $L=256$. We can clearly observe that the
best condition for shortest path length is obtained for $\alpha=3$.}
\label{2}
\end{figure}

Optimal navigation with the presence of long-range links in a lattice
network without constraints was studied recently by
Kleinberg~\cite{Klein2000a}. Here we show for the first time that a
rather different behavior can be observed for this problem when
realistic constraints on total length are imposed on the process of
adding long-range connections. To better demonstrate the competition
between total length and number of links, we generate a single network
realization ($L=256$) for a given value of $\alpha$ and compute the
shortest-path length, $\ell$, from each node in the network to its
central node. This calculation is performed as follows. Once we choose
the root node (e.g., the central one), we visit all its neighbors,
including the neighboring nodes connected by long-range
connections. These visited nodes are classified as {\it shell one
nodes}, meaning that they are only one time step away from the root
node~\cite{FN3}. After that, we visit all the neighbors of these nodes
not visited before and classify them as {\it shell two
nodes}. Following this procedure for all network nodes, we obtain the
$\ell$ values (time) for each node to be reached from the root
node. Figure~\ref{2} shows the contour plots representation of the
$\ell$ values performed for four different values of the parameter
$\alpha$. For $\alpha=1$ and $2$ the number of long-range connections
is small. As a result, only a few little islands sparsely dispersed in
these networks are really close to their central nodes (only a few
short and/or long-range connections away). For $\alpha=3$ the added
long-range links are shorter, but more numerous, thus substantially
decreasing the shortest path over the whole network. For $\alpha=4$,
due to the very short length size of the added connections, only a
limited region surrounding the central node displays a reduced
shortest path. Sites which are further away from the origin have
significantly larger path length $\ell$ to the origin.

\begin{figure}[t]
\begin{center}
\includegraphics*[width=8cm]{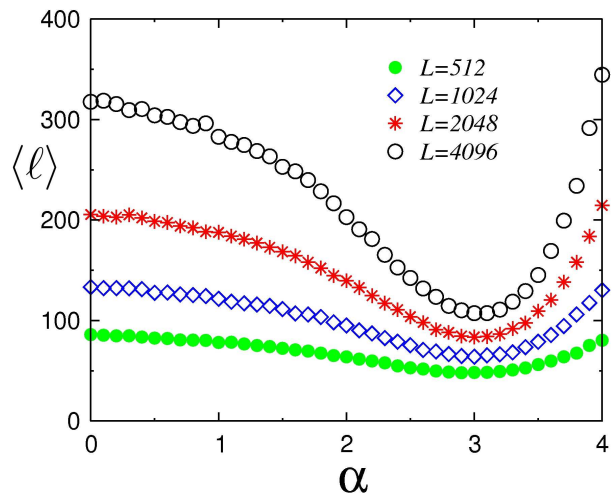}
\end{center}
\caption{(Color online) Average shortest path length $\av{\ell}$ as a 
function of $\alpha$. There is a constraint in the total length of the
long-range connections, $\Lambda=\sum r_{ij}=L^{2}$, where $L$ is the
size of the underlying square lattice.  We find that the optimal
shortest-path is achieved for $\alpha=3$. With the restriction on
total length, the number of long-range connections are not fixed
(e.g.. with $\alpha=0$, large long-range connections become frequent,
which reduces the total number of long-range connections.) To obtain
these results, we simulated $10,000$ realizations for $L=512$, $3500$
realizations for $L=1024$ and $2048$, and $25$ realizations for
$L=4096$.}
\label{3}
\end{figure}

We extract more quantitative information about this navigation problem
by performing extensive simulations for different values of $\alpha$
and many realizations of different system sizes. In each case, the
average shortest path $\av{\ell}$ is calculated over all realizations,
considering all the shortest distances between each pair of nodes. We
assume that the total length (cost) is proportional to the total
length of the links in the underlying network, i.e., $\Lambda=AL^{2}$,
where $A$ is a constant. That is, the budget to improve the system is a
fraction of the cost of the current network (without long-range
connections)~\cite{FN2}.

\begin{figure}[t]
\begin{center}
\includegraphics*[width=8.5cm]{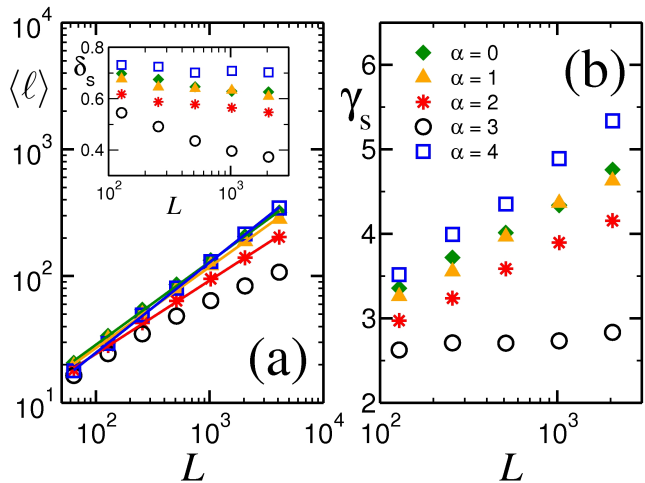}
\end{center}
\caption{(Color online) In (a) we show the average shortest path length 
$\av{\ell}$ as a function of the lattice size $L$ for the square
lattice. The constraint in the total length of the long-range
connections is $\Lambda=L^{2}$. The curve with $\alpha=3$ increases
slower with $L$ compared to any other value of $\alpha$. In the inset,
the plot of the successive slopes $\delta_{S}$ obtained from
$\log_{10}\av{\ell}$~versus~$\log_{10}L$ reinforces the display of
power-law behavior of $\av{\ell}$ with $L$ for $\alpha \neq 3$. The
plot of the successive slopes $\gamma_{S}$ obtained from
$\log_{10}\av{\ell}$~versus~$\log_{10}(\log_{10}L)$ shown in (b)
indicates that $\av{\ell}$ increases as a power of the logarithm of
$L$ for the optimal condition $\alpha=3$.}
\label{4}
\end{figure}

The results presented in Fig.~\ref{3} clearly indicate the presence of
a minimum $\av{\ell}$ for different system sizes at the same value of
the exponent $\alpha=3$, where optimal navigation is achieved with
(global) knowledge of the shortest paths and cost limitations. The way
in which $\av{\ell}$ scales with system size $L$, however, seems to
follow rather different behaviors, depending on the value of
$\alpha$. We tested two possible forms for $\av{\ell}$ vs $N$, a power-law
and a power of $\log N$. As shown in the main plot of Fig.~\ref{4}a,
our results for $\alpha \neq 3$ suggest that the shortest path
$\av{\ell}$ follows a power-law with system size $L$.  This is
supported by the plot of successive slopes $\delta_{S}$ obtained from
$\log_{10}\av{\ell}$~versus~$\log_{10}L$, which are almost invariant,
as shown in the inset of Fig.~\ref{4}a. In contrast, in the case of
$\alpha=3$, the increase with $L$ of $\av{\ell}$ appears to be less
rapid than a power-law. Interestingly, the successive slopes $\gamma_{S}$ 
obtained from $\log_{10}\av{\ell}$~versus~$\log_{10}(\log_{10}L)$, as 
presented in Fig.~\ref{4}b, indicate that $\av{\ell}$ increases as a 
power of the logarithm of $L$, $\av{\ell}\sim \log_{10}^{\gamma_{S}}L$, 
rather than a power of $L$, only for $\alpha=3$. This provides clear
support for the fact that in the optimal condition, $\alpha =3$, the
transport will improve even further as $L$ increases, as suggested by
Fig.~\ref{3}.

We also studied our model for a one-dimensional lattice and observed
similar behavior. The optimal condition we obtained in this case at
$\alpha=2$ (data not shown) leads us to conjecture that the optimal
value is obtained at $\alpha=d+1$, where $d$ is the dimension of the
underling lattice. Note that the Kleinberg result was extended to
fractals~\cite{Rober2006}, where the optimal exponent is found to be
$\alpha = d_{f}$, namely, the fractal dimension of the substrate. The
$\av{\ell}$ dependence on $L$ for different $\alpha$ for the Kleinberg
model was recently derived analytically~\cite{Carmi2009}.

In the following, we present analytical arguments showing that $\alpha
=3$ is indeed the only case where logarithmic scaling of $\av{\ell}$
with $L$ can occur, while for $\alpha \neq 3$ a power-law with $L$
should exist. By arbitrarily fixing the cost parameter to
$\Lambda=AL^{2}$, we obtain that $\rho\sim {\av{r}}^{-1}$, where
$\rho$ is the density and $\av{r}$ is the average length of the added
long-range connections. Since $\av{r} \sim \int_{1}^{L}
r^{2-\alpha}dr$, it follows that for $2 \leq \alpha < 3$, $\rho \sim
L^{\alpha-3}$ and for $\alpha<2$, $\av{r}$ is limited by the network
size leading to $\rho\sim L^{-1}$. Thus, for all values of $\alpha <3$
the density of the long-range links added, due to the constraint,
decreases as a power-law with $L$. As a consequence of this power-law
decrease in density, $\av{\ell}$ must increase as a power of $L$. To
see this we argue that $\av{\ell}$ is bounded by the relation
$\av{\ell} > \rho^{-1/d}$. The right hand side, $\rho^{-1/d}$, appears
for the case of the small world model, where $\alpha=0$, with a fixed
concentration of links, $\av{\ell}\sim \rho^{-1/d}\ln
L$~\cite{Barth1999}. Since for the case $0<\alpha<3$, $\av{\ell}$
decreases with increasing $\alpha$, the bound $\av{\ell} >
L^{(3-\alpha)/d}$ is rigorous and $\av{\ell}$ in this range must scale
as a power of $L$. For $\alpha>3$ and sufficiently large networks,
$\av{r}$ is finite and the density becomes independent of the system
size, i.e., $\rho \sim L^{0}$. Thus, the effect of the constraint
$\Lambda$ on navigation should become negligible. However, the finite
value of $\av{r}$ suggests that long-range links can be neglected and
therefore $\av{\ell}$ should scale as a power of $L$. Thus, it follows
that only for $\alpha =3$, $\av{\ell}$ can scale logarithmically with
$L$, as suggested by our numerical simulations (see Fig.~\ref{4}).

\begin{figure}[t]
\begin{center}
\includegraphics*[width=8.5cm]{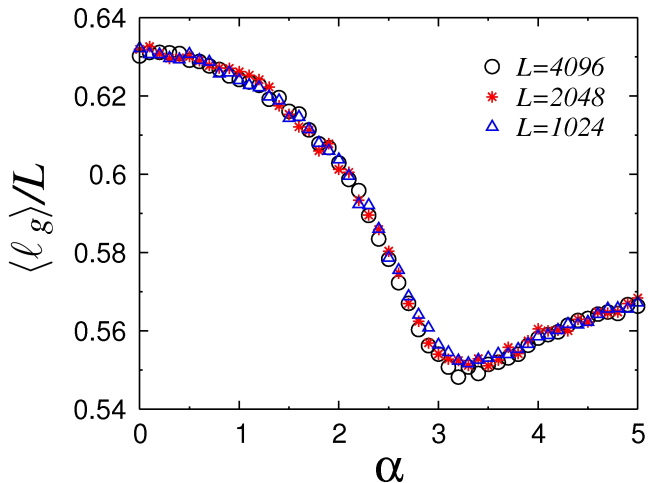}
\end{center}
\caption{(Color online) The characteristic average delivery time 
$\av{\ell_g}/L$ as a function of $\alpha$ for navigation with the {\it
greedy algorithm}. The cost $\Lambda$ involved to add long-range
connections changes the behavior of the density of long-range
connections. As a result of that, a minimum is observed at
$\alpha\approx 3$. Each data point is a result of $4000$ simulations and
the cost $\Lambda$ is fixed at $L^{2}$.}
\label{5}
\end{figure}

It is important to note that our global navigation scheme with
$\av{\ell}$ can be considered as a lower bound to any other transport
navigation process. For example, a strategy based on purely local
knowledge of the network structure will necessarily perform worse than
any other with global information. In Ref.~\cite{Klein2000a}, for
example, the {\it greedy algorithm} is introduced as a paradigm based on
local information, where the traveler, when leaving a node, chooses to
move to the one among its neighbors which has the smallest Manhattan
distance to the target. Kleinberg found that $\alpha=2$ is the optimal
value in the navigation with the greedy algorithm~\cite{Klein2000a}. We
next ask, what would be the optimal $\alpha$ for the greedy algorithm
when cost restriction $\Lambda=AL^{2}$ is imposed? We find also for the
greedy algorithm that the optimal value is $\alpha=3$. This is shown in
Fig.~\ref{5}, where we plot the average delivery length $\av{\ell_{g}}$
that a message travels with only local information of the system
geometry. The message is sent from the source node $s$ to the target
node $t$ through a network generated with the constraint
$\Lambda=L^{2}$. Remarkably, the presence of a minimum also at
$\alpha \approx 3$ shows that the type of information (local or global)
used by the message holder to pass it through the system during the
navigation process becomes unimportant if the network is constructed
under length (cost) limitations.  However, the two mechanisms display
very different and distinct behaviors regarding the scaling with system
size. While we observe logarithmic growth for the optimal condition
$\alpha=3$, in the case of global information, the time to reach the
source, with the greedy algorithm and with cost constraint, appears to
increase linearly with size for all values of $\alpha$. The linearity of
$\av{\ell_g}$ with $L$ is observed in the scaling collapse (Fig.~\ref{5})
of the curves of $\av{\ell_g}/L$ vs. $\alpha$.

In summary, we have investigated the effect of introducing a cost
constraint in the optimal design of a transportation network. Our
results show that, regardless of the strategy used by the traveler,
based on local or global knowledge of the network structure, the best
transportation condition is obtained with an exponent $\alpha=d+1$,
where $d$ is the topological dimension of the underlying lattice. The
results recently reported by Bianconi {\it et al.}~\cite{Bianc2008} on
the US airport network provide striking support for the validity of our
optimal exponent $\alpha=3$. The fact that the probability of a flight
connection within US decays as a power-law with the distance between
airports, $r^{-\alpha}$, where $\alpha=3.0 \pm 0.2$, reveals the
optimized aspect of the network under the conditions of uniform
geographical availability (for customer satisfaction) and cost
limitations (for airline companies profit). The result $\alpha=3$ is in
sharp contrast with the results obtained for unconstrained systems with
global and local information, where the optimal conditions are
$\alpha=0$~\cite{Komid2008} and $\alpha=d$~\cite{Klein2000a,Rober2006},
respectively.  The contrast between the optimal results is even more
dramatic. While in the unconstrained case the mean length of a link
{\it diverges}, we find that when cost is considered the mean length is
{\it finite}. In the case where the traveler has global knowledge of the
network, and is able to identify the shortest path for navigation, we
obtain a slow (logarithmic) growth with size for the transit time at the
optimal condition. A different picture is obtained if the traveler has
only local knowledge of the network. For example, in the case where the
transportation path is decided based on the Manhattan distance to the
target, we obtain a linear growth of the transit time with system size,
for all values of the exponent $\alpha$. Finally, our results suggest
that the idea of introducing a cost constraint in the navigation problem
offers a different and more realistic theoretical framework to
understand the evolving topologies of other important complex network
structures in nature, such as subways, trains, or the Internet.

We thank CNPq, CAPES, FUNCAP, FINEP, ONR, the Israel Science Foundation,
and the European Project EPIWORK for financial support, and
D. ben-Avraham for useful discussions.

%%%%%%%%%%%%%%%%%%%%%%%%%%%%%%%%%%%%%%%%%%%%%%%%%%%%%%%%%%%%%%%%%%%%%%%%%%%%


\begin{thebibliography}{99}
\bibitem{Alber1999}
R. Albert and A.-L. Barab\'asi, Nature (London) \textbf{401}, 130
(1999).

\bibitem{Watts1998}
D. Watts and S. Strogatz, Nature \textbf{393}, 440 (1998).

\bibitem{Barth1999}
M. Barth\'el\'emy and L. A. N. Amaral, Phys. Rev. Lett. \textbf{82},
3180, 5180 (1999).

\bibitem{Jeong2000}
H. Jeong et al., 
%B. Tombor, R. Albert, Z.N. Oltvai and L.-A. Barab\'asi,
Nature (London) \textbf{407}, 651 (2000).

\bibitem{Lawre1999}
S. Lawrence and C. L. Giles, Nature (London) \textbf{400}, 107
(1999).

\bibitem{Giles1998}
C. L. Giles et al., 
%S. Lawrence and B. Krovretz, 
Science \textbf{280}, 1815 (1998).

\bibitem{Klein2000a}
J. M. Kleinberg, Nature (London) \textbf{406}, 845 (2000); Proc. 32nd
ACM Symposium on Theory of Computing 163--170 (2000).

\bibitem{Rober2006}
M.R. Roberson and D. ben-Avraham, Phys. Rev. E \textbf{74}, 017101
(2006).

\bibitem{Carmi2009}
S. Carmi et al., 
%S. Carter, J. Sun, and D. ben-Avraham, 
Phys. Rev. Lett. \textbf{102}, 238702 (2009).

\bibitem{Cart2009}
C. C. Cartozo and P. De Los Rios, Phys. Rev. Lett.
\textbf{102}, 238703 (2009).

\bibitem{Komid2008}
K. Komidis et al., 
%S. Havlin and A. Bunde, 
Europhys. Lett. \textbf{82}, 48005 (2008).

\bibitem{Mouka2002}
C.F. Moukarzel, and M. A. de Menezes, Phys. Rev. E \textbf{65}, 056709
(2002).

\bibitem{Guim2002} 
R. Guimera et al., 
%A. Diaz-Guilera, F. Vega-Redondo, A. Cabrales, and A. Arenas, 
Phys. Rev. Lett. \textbf{89}, 248701 (2002).

\bibitem{Danon2008}
L. Danon et al., 
%A. Arenas, A. Diaz-Guilera, 
Phys. Rev. E \textbf{77}, 036103 (2008).

\bibitem{Santos2008} 
M. C. Santos et al., 
%G.M. Viswanathan, E.P. Raposo, M.G.E. da Luz,
Phys. Rev. E \textbf{77}, 041101 (2008).

\bibitem{Boguna2009} 
M. Boguna and D. Krioukov, Phys. Rev. Lett. \textbf{102}, 058701
(2009).

\bibitem{Lator2002}
V. Latora and M. Marchiori, Physica A \textbf{314}, 109 (2002).

\bibitem{Youn2008}
H. Youn et al., 
%M.T. Gastner, and H. Jeong, 
Phys. Rev. Lett. \textbf{101}, 128701 (2008).

\bibitem{FN3}
We assume that the traverse time of a link is almost the same for both
short-range and long-range links since most of the time is spent in
station and for decelerating and accelerating.

\bibitem{FN2}
Here, we show the case $A=1$ but we obtained similar results for several
values of $A$, $0<A<1$.

\bibitem{Bianc2008}
G. Bianconi et al., ``How relevant are features for network structure?'',
%P. Pin, and M. Marsili, 
arXiv:0810.4412 (2008).

\end{thebibliography}
\end{document}